\documentclass [11pt]{article}
\usepackage {amssymb}
\usepackage {amsmath}
\usepackage {graphicx}
\def\be{\begin{equation}}
\def\ee{\end{equation}}
\def\vp{\vspace{.20in}}

\begin{document}
\renewcommand{\baselinestretch}{1.1}
\vp
\vp
\Large
\centerline{Opportunistic infection as a cause of transient viremia in }
\centerline{chronically infected HIV patients under treatment with HAART} 
\normalsize 
\vp    
\vp     
\centerline{Laura E. Jones }  
\centerline{\textrm{Ecology and Evolutionary Biology}}
\centerline{\textsc{Cornell University}}
\centerline{Ithaca, New York 14853}
\vp
\centerline{and}
\vp
\centerline{Alan S. Perelson}  
\centerline{\textrm{Theoretical Biology and Biophysics}}
\centerline{\textsc{Los Alamos National Laboratory}}
\centerline{Los Alamos, New Mexico 87545}
\vp
\pagebreak

\section*{Abstract}

When highly active antiretroviral therapy is administered
for long periods of time to HIV-1 infected patients, most patients achieve
viral loads that are ``undetectable'' by standard assay 
(i.e., HIV-1 RNA $ < 50$ copies/ml).  Yet despite exhibiting sustained 
viral loads below the level of detection, a number of 
these patients experience unexplained episodes of transient viremia
or viral ``blips".  We propose here that transient activation of
the immune system by opportunistic infection may explain these
episodes of viremia. Indeed, 
immune activation by opportunistic infection may spur HIV replication, replenish viral reservoirs
and contribute to accelerated disease progression.
In order to investigate 
the effects of concurrent infection on chronically infected HIV patients 
under treatment with highly active antiretroviral therapy (HAART), 
we extend a simple dynamic model of the effects
of vaccination on HIV infection [Jones and Perelson, JAIDS 31:369-377, 2002] 
to include growing pathogens. We then propose a more realistic model 
for immune cell expansion in the presence of pathogen, and include this
in a set of competing models that allow low baseline viral loads in the presence of drug treatment.
Programmed expansion of immune
cells upon exposure to antigen
is a feature not previously included in HIV models,
and one that is especially important to consider when simulating an immune response to
opportunistic infection.
Using these models we show that viral blips with realistic duration and amplitude can be
generated by concurrent infections in HAART treated patients.

\section*{Introduction}

\begin{figure}
\centerline{\includegraphics[width=15.25cm,height=7.62cm]{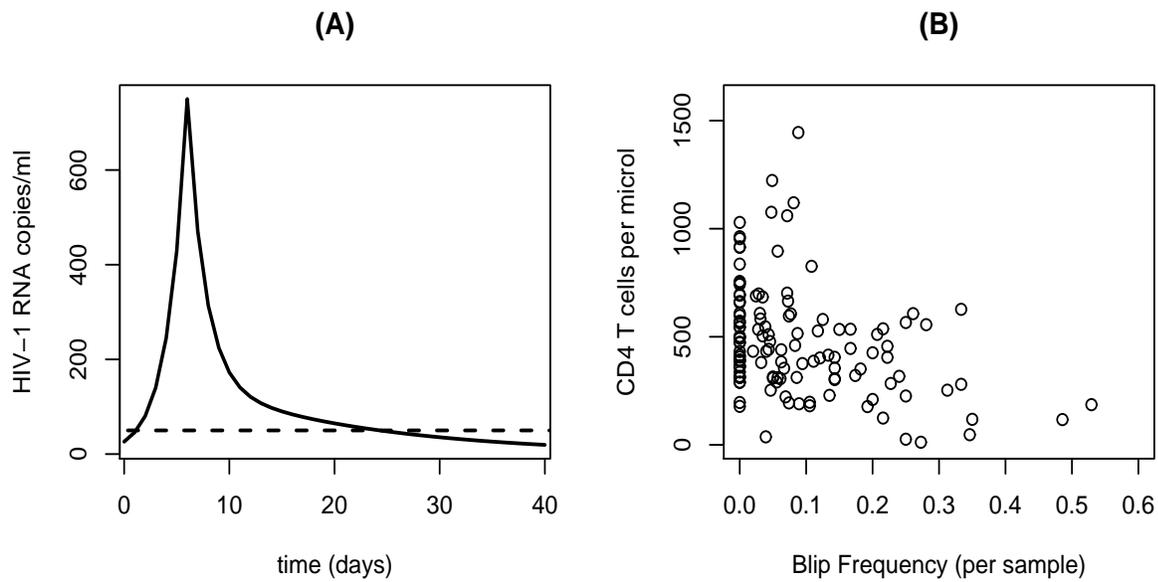}}
\caption{\small Generalized viral blip shape, and correlation with CD4+ T-cell count
at the onset of therapy. A, stylized blip with rapid rise and two-phase exponential
decay. Given a mean duration over
threshold of 20 to 30 days, the two decay constants assume values of roughly $0.6$ 
and $0.06 d^{-1}$ (Di Mascio et al. 2003a), consistent with the mortalities of 
short lived productively infected cells and longer-lived chronically infected
cells. B, CD4+ count at the start of therapy versus viral blip frequency. 
Both panels after Figure 4 (panels B,C), Di Mascio {\it et al.} 2003a.
}
\label{BlipsFigure}
\end{figure}

Adherence to a regime of highly active antiretroviral therapy (HAART)
suppresses the viral loads of most chronically infected
HIV patients below the level of detection by standard assay.  However,
a number of these otherwise well-suppressed patients experience 
unexplained ``viral blips"  while on therapy.
Di Mascio et al. (2003a) in a study of  123 patients  found that
the mean blip amplitude was $158 \pm 132$ HIV RNA copies/ml, with
the distribution skewed towards low amplitude blips. It addition,
Di Mascio et al. (2003a)  suggest that a viral blip is not an isolated event
but rather is a transient, 
intermittant episode of detectable viremia (HIV-1 RNA $> 50$ copies/ml)
with a duration of roughly two to three weeks  
They further showed that
the amplitude and frequency distribution 
of these viral transients are consistent with viral load rising sharply,
followed by slower, two-phase (double exponential) decay (Di Mascio et al., 2003a).
An example of a typical blip is shown in Figure \ref{BlipsFigure}~A.
The frequency of these episodes appears to
be inversely correlated with CD4+ T-cell count at baseline, prior
to initiation of drug therapy
(Di Mascio et al., 2003a, 2004b).
Figure \ref{BlipsFigure}~B shows data from the 123 patients in the
Di Mascio et al. 2003a study.  CD4+ T cell counts at the
onset of therapy are plotted against the frequency of blip occurances per sample, showing
a positive correlation between low initial CD4+ counts and blip frequency.

A number of possible causes of blips have
been suggested, including  but not limited to missed drug doses,
activation of latently infected cells and consequent release 
of virus, release of virus from tissue reservoirs, and a rise in target cell availability 
due either to  vaccination (Jones and Perelson, 2002) or  one or more coinfections
by opportunistic pathogens, which then increase viral replication.  The observation
of Di Mascio et al. (2003a) that blip frequency is inversely correlated with
baseline CD4+ T cell count suggests that
patient specific factors, such as susceptibility to infection,
which increases at low CD4 counts, may play a role in blip generation.
 Prior work on untreated, chronically infected
HIV patients documents increases in viral load associated with vaccination
(O'Brien et al., 1995; Staprans et al., 1995; Brichacek et al., 1996; 
Stanley et al., 1996) and with opportunistic infection (e.g., Donovan et al., 1996).
McLean and Nowak (1992) proposed some of the first models of enhanced
HIV replication due to immune stimulation via opportunistic infection, and
showed how the positive feedback between enhanced HIV replication and incomplete
immune control of pathogens due to HIV--immunosuppression leads finally to
immune failure and full--blown AIDS.

This paper examines the hypothesis that viral blips result from random encounters with replicating antigens - or transient opportunistic infections in HIV positive patients. We begin with a simple model for
coinfection, explore its biological shortcomings, and develop a series of incrementally
more complex models. To minimize the introduction of new parameters while accurately simulating the dynamics of transient viremia, we add missing biology at each step and then test each incremental model to see where it falls short and where it is sufficient. In the process, we introduce  biologically realistic mechanisms for programmed immune cell proliferation, and include features which allow
robust low viral loads under drug treatment.  

\section*{A simple model for coinfection}
We begin with the following simple model for coinfection in the presence
of HIV infection, which is a generalization of the HIV infection model developed
by Perelson et al. (1996, 1997) and reviewed by Perelson (2002).

\begin{subequations} 
\label{eqs1}
\begin{eqnarray}
\frac{dA}{dt} &  = &r_0A(1-\frac{A}{A_{max}})  - \gamma AT \\ 
\frac{dT}{dt} &  = & \lambda + a(\frac{A}{A + K})T - d_T T - (1-\epsilon) kVT  \\
\frac{dT^*}{dt} & =  & (1 - \alpha)(1-\epsilon) kVT - \delta T^*  \\
\frac{dC}{dt} & = & \alpha(1-\epsilon) kVT - \mu C \\
\frac{dV}{dt} & = &  N_T\delta T^* + N_c\mu C - cV  
\end{eqnarray}
\end{subequations}

Here the antigen, $A$, is a growing pathogen,  $T$ are uninfected CD4+ T 
cells, $T^*$ are cells productively infected with HIV, $C$ are cells
chronically infected with HIV, and $V$ represents HIV-1 (RNA copies/ml).  Pathogen $A$
undergoes density-dependent growth described by a logistic law with maximum growth
rate $r_0$ and carrying capacity $A_{max}$. As in earlier work
by Jones and Perelson (2002), we assume that the antigen
is cleared in a T cell-dependent manner with rate
constant $\gamma$.  We further assume that
uninfected T cells, $T$, are generated at rate $\lambda$,
die at rate $d_T$, and are infected by virus with rate constant 
$k$. Assuming reverse transcriptase (RT) inhibitors are administered,  the
infectiousness of the virus $k$ is reduced by $(1-\epsilon)$, where $\epsilon$ is
the efficacy of the RT inhibitors and $0 \leq \epsilon \leq 1$.

In this simple model we assume T cells are activated into
proliferation at a maximum rate $a$ in the presence of pathogen, 
and that the proliferation rate depends on the pathogen 
concentration with a half-saturation constant $K$. 
HIV infection of T cells results in 
productively infected cells $T^*$, which die at a rate
$\delta$,  and  chronically
infected cells $C$, with mortality $\mu < \delta$. A fraction of infection events $\alpha << 1$
results in chronic infection. Chronically infected cells live much longer, producing virus more slowly
than productively infected cells. The inclusion of the chronically infected pool is motivated by the
suggested two-phase decay of a viral transient (Di Mascio et al., 2003a).

Finally, virus is  produced by productively and chronically infected cells at  
rates $N_T\delta$ and $N_C\mu$, respectively, where $N_T$ and $N_c$ are average burst sizes
for productively and chronically infected cells. Virus is cleared at a constant rate $c$ per
virion.

Based on previous work we take as a typical set of parameter values $\lambda = 1 \times 10^4$ ml$^{-1}$,
$k=8 \times 10^{-7}$ ml/d, $\alpha=0.195$, $N_T=100$ and $N_C=4.11$ (Callaway and Perelson, 2002);
$d=0.01$ d$^{-1}$ (Mohri et al., 1998); $\delta=0.70$ d$^{-1}$ and $\mu =0.07$ d$^{-1}$ (Perelson et al., 1997);
and $c=23$ d$^{-1}$ (Ramratnam et al., 1999). 
The antigen or pathogen clearance rate 
constant $\gamma$ is a ``fitted'' parameter, which we set to $1 \times 10^{-3}$ ml d$^{-1}$.  
In a prior study of the effects on viral load of vaccination (with a non-replicating antigen), 
this fitted value varied widely from patient to patient with values
roughly $1 \times 10^{-5}$ to $1 \times 10^{-8}$ ml d$^{-1}$,
reflecting differences in patient immune response (Jones and Perelson, 2002). 
We assumed a higher clearance rate in this study since immune activation by a proliferating
pathogen should be much greater than activation due to a small dose of non-replicating antigen.
Note that where these parameters are used elsewhere in this paper, they retain the above values
unless otherwise noted. 

In the absence of pathogen, and assuming chronic HIV infection, the viral load
stabilizes at the following equilibrium state:

\be
\label{ss1}
\bar V = \frac{\lambda}{c}[ (1-\alpha)N_T + \alpha N_c ] - \frac{d}{(1-\epsilon) k} \ ,
\ee

implying an inverse relationship between steady state viral load and
drug efficacy $\epsilon$.

This model without the inclusion of drug therapy, i.e. $\epsilon=0$, and with 
a non-growing pathogen, i.e., $r_0=0$,
 worked surprisingly well for modeling
HIV dynamics in chronically infected, untreated patients
following vaccination with a common recall antigen
(Jones and Perelson, 2002). 

However, this model does not
generate realistic blips when a growing pathogen is substituted for a
vaccine. There are several reasons for this: even in untreated patients,
according to our model, T-cells respond very rapidly, eliminating the pathogen
before it has time to grow (Figure \ref{Figure1abc}~A), so there is only a relatively
small immune response to the presence of the pathogen, and little change in
viral load (Figure \ref{Figure1abc}~B).  The system remains relatively unaffected by increases in pathogen growth rate 
until $r_0$ reaches a critical point where the model immune system cannot completely
eliminate the pathogen, and then there are predator-prey cycles. 
[Note that inclusion of a logistic growth
term means that the pathogen can reach a carrying capacity, $A_{max}$, but will not
experience runaway growth.]  With the addition of drug therapy (Figure \ref{Figure1abc}~C) 
of efficacy $\epsilon=0.6461275$ (see Appendix), baseline viral load is suppressed to 25 RNA copies (ml$^{-1}$)
and opportunistic infection results in a small, slow rise and fall in viral load, rather than a 
burst of viremia.   If baseline viral load is
suppressed further, then there is no appreciable change in viral load.

\section*{Building a new model}

While different parameters could be explored, we believe that there is a need
to generate more biologically realistic
models.  First, the majority of patients with viral loads below 50 copies/ml who have
 been examined with more sensitive assays have viral loads that are still detectable, 
i.e. been 1 and 50 copies/ml (Dornadula et al.,1999, Di Mascio, 2003b). This sugests that appropriate models must exhibit robust low viral steady states for patients on therapy. 
By contrast, the model given by
eqs. (\ref{eqs1}) requires drug efficacy $\epsilon$  precisely tuned to many decimal places 
in order to yield a low, yet not vanishingly small,  viral steady state 
(Callaway and Perelson, 2002).  Second, recent immunological
data suggest T-cell proliferation stimulated by a pathogen involves a cascade of divisions,
all triggered by a brief exposure to antigen.  Models should thus account for this
type of ``programmed" response.  Lastly,  immune response to a pathogen involves 
generation of effector cells that are responsible for clearing the pathogen. For many
viral infections, CD8+ T cells are critical for containing and clearing the
infection (Wong and Palmer, 2003).  This is well documented in lymphocytic choriomeningitis
 virus (LCMV) infection, where a
strong antigen-specific CD8+ T cell response is induced, leading to rapid elimination of the virus
(Murali-Krishna et al., 1998a,b).

\begin{figure}
\centerline{\includegraphics[width=15.25cm,height=7.62cm]{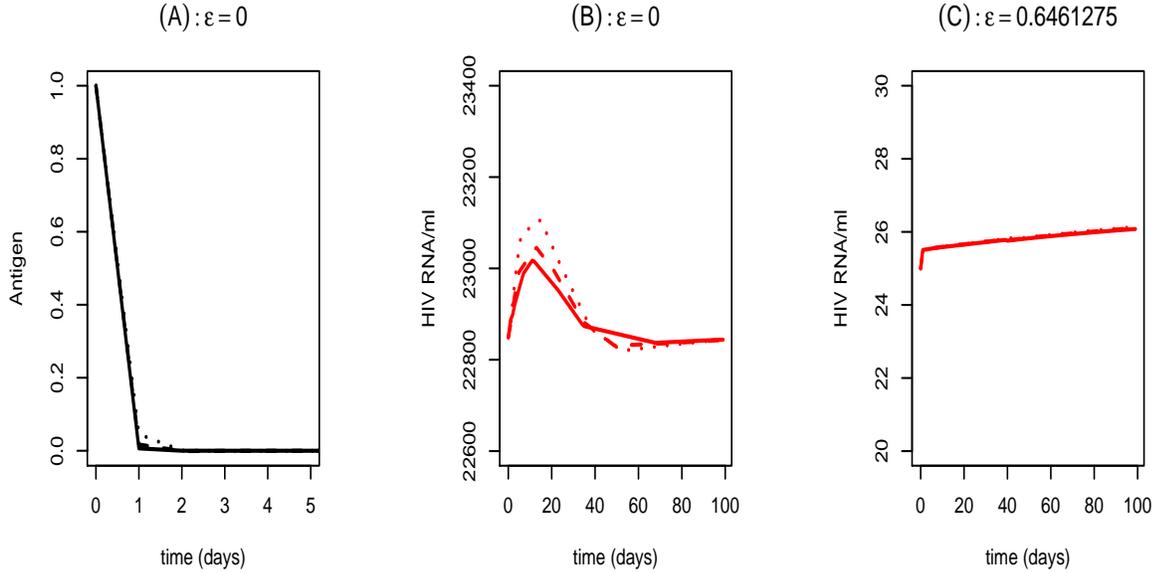}}
\caption{\small \small Simulations of a simple coinfection model (\ref{eqs1}) for
pathogen growth rates $r_0 = 1,2,3,4$ day$^{-1}$ (solid, dashed, dotted line in each panel) given
no drug treatment (panels A, B) and treatment with a drug of efficacy  $\epsilon=0.6451275$ (panel C).
A, Antigen growth assuming $\epsilon=0$ and $A_0 = 1$; B Viral load (HIV-1 RNA/ml) assuming $\epsilon=0$;
C, Viral load (HIV-1 RNA/ml) assuming  $\epsilon=0.6461275$. Baseline VL is 25 RNA copies/ml. }
\label{Fig1}
\end{figure}

\subsection*{T-cell proliferation under antigenic stimulation} 

Upon exposure to antigen, naive CD8+ T cells undergo a burst of proliferation,
entering a programmed cascade of divisions that culminate in the production of
mature, activated effector cells (Kaech and Ahmed, 2001).  This is followed by a programmed contraction in
which most of the effector cells are subject to apoptosis, leaving a small, stable
memory population (Badovinac et al, 2002).  Revy et al. (2001) proposed a system of ordinary differential 
equations to analyze and describe T-cell proliferation.
We adapt their model to incorporate activation by a growing pathogen, $A$, which 
is finally cleared by CD8+ effector cells, $E$.  

\begin{subequations}
\label{eqs2}
\begin{eqnarray}
\hbox{Antigen}\quad \frac{dA}{dt} & = & r_0A(1- \frac{A}{A_{max}}) - \gamma AE \\
\frac{dN_0}{dt} &  = & -( p_0(A) + d_0)N_0 \\
\hbox{CD8 Response}\quad \frac{dN_1}{dt} &  = & 2  p_0(A)N_0 - (p + d)N_1 \\
\frac{dN_{i}}{dt} & = &\begin{cases}  2p N_{i-1} - (p + d)N_{i},   i=2,3,4. \cr
                      2p N_{i-1} - (p + d_E)N_{i},   i=5,..., k-1. \cr \end{cases} \\
\frac{dN_k}{dt} & = & 2p N_{k-1} - d_E N_k  
\end{eqnarray}
\end{subequations}

Here  again $A$ is an opportunistic pathogen, $r_0$ is the pathogen
growth rate, and  $\gamma$ is the clearance rate constant  for the pathogen.
$p_0$ and $p$ are constant proliferation rates, $N_0$ is the
initial, naive cell pool, and the $N_i$ are proliferative phases, or
``division classes'': for each $i$, the number of cells that have completed $i$ divisions, and
$E$ are mature, pathogen-specific ``effector" cells. 
As the CD8+ T cells proliferate they also differentiate into effectors, $E$. Here we assume
 cells become effectors after 4 divisions and stop proliferating after
8 division; $E = \sum_{i=4}^8 N_i$. Alternatively,
one could assume a fraction of cells, $\beta_i$, become effectors after $i$  divisions,
i.e., $E= \sum_{i=1}^8 \beta_i N_i$, or leave the maximum number of divisons as
an adjustable parameter. 
Proliferative, non-effector phases (division classes $N_0$, $N_1$, ..., $N_3$)
undergo mortality at a rate $d << d_E$, the death rate for
activated effector cells.  

Experiments suggest that when quiescent cells are stimulated into
proliferation, the initial cellular division takes longer than subsequent
divisions (Gett and Hodgkin, 2000), and that the time to first division depends
on features of the antigen stimulation.  For example, using
anti-CD3 antibodies rather than antigen to stimulate CD8+ T cells,
Deenick et al. (2003) found that decreasing the anti-CD3 concentration lengthened 
the time to first division.  
We thus assume the rate of the first
CD8+ T cell division depends on antigen according to a
Type-III functional response, 

\be
 p_0(A) = p_0\frac{A^n}{(A^n + K_8^n)}
\label{fA}
\ee
 
where $K_8$ represents the critical antigen level required to stimulate a
CD8+ response [Table I]. Thus when $A$ is low, i.e., $A << K_8$, growth is slow,
whereas when $A$ attains values greater than $K_8$ rapid growth ensues, saturating at 
rate $p_0$.  Here $n$ is a parameter frequently called the Hill coefficient and it
 determines the steepness of the  response when $A$ is near the ``threshold" $K_8$.

While the initial division may be relatively slow,  subsequent divisions occur more
rapidly, without an initial delay, as cells are already activated. These divisions
occur even if antigen is removed (Badovinac et al., 2002). We 
therefore assume subsequent divisions occur at a constant rate, $p>p_0$ (Table 1).

In the model we do not explicitly account for precursor cells that fail to divide, although 
in the experiments by Gett and Hodgkin (2002) such cells are present.

Based on CD8+ T-cell response estimates for LCMV in mice (De Boer et al., 2001),
we set  $p = 2.92$ d$^{-1}$, and chose the initial activation/first division rate
$p_0 =  1$ d$^{-1}$. The initial quiescent cell
pool is small, $N_0 = 500$ cells/ml: it is estimated that one in $10^5$ T cells is specific for any given
epitope, and assuming that a healthy human being has roughly $10^{11}$ CD8+ T cells,
we expect about $10^6$ antigen-specific cells. 
Most naive cells are in lymphoid tissue, which in a 70 kg man weighs about 700 grams.
Thus, with some cells circulating in the blood and others not naive, a density of about
500 cells/ml in lymphoid tissue, where most proliferation occurs, seems appropriate.

\begin{figure}
\centerline{\includegraphics[width=15.25cm,height=7cm]{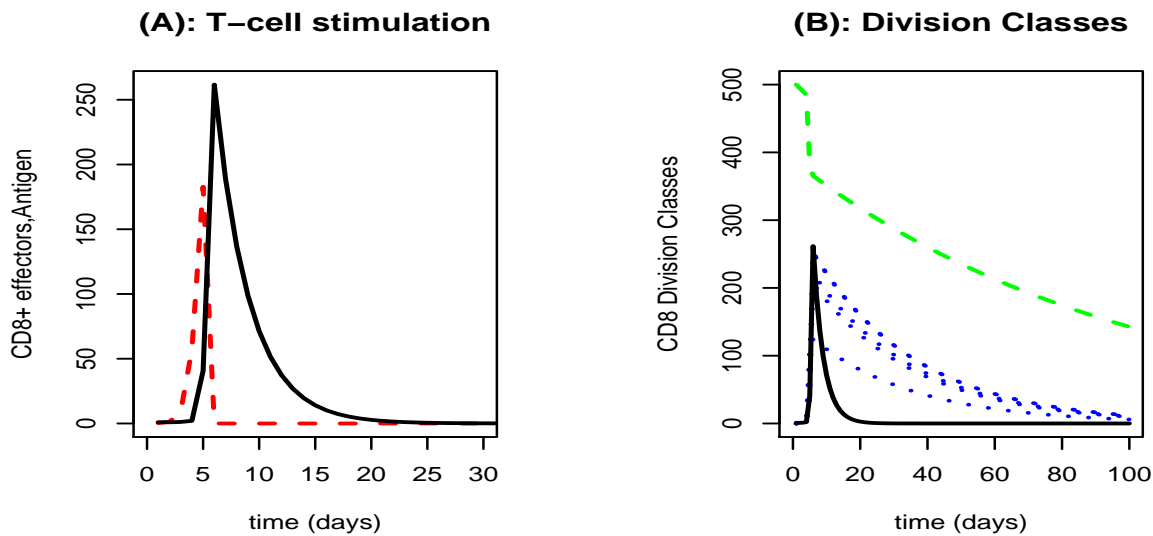}}
\caption{\small Simulated results for CD8+ T-cell proliferation stimulated by a growing
antigen. A, antigen ($A_0 =1$), shown in dashed red line; 
total CD8+ effector cells (solid black line). B, cell counts for the various division classes
(dotted blue);
total effectors  ($\sum_{i=4}^8 N_i$) (solid black line), and naive cells (dashed green line).
 Since the process of proliferation is a cascade involving both growth and mortality and we assume
growth is more rapid than mortality
each successive division class is slightly 
larger than the preceding class. The effector population (black line) is subject to higher mortality
than the less differentiated preceeding classes.}
\label{Figure 2}
\end{figure}

Figure \ref{Figure 2} shows results from simulation of equations (\ref{eqs2}). 
In this simulation, we assume eight division classes, $N_i,\ i=1,...,8$,
plus naive cells, $N_0$, with effector cells, $E$, corresponding to the sum of the
last four division classes.
In panel A, the growing antigen (dashed red line), and CD8+ effectors (solid black line)
are shown. 
Panel B shows the CD8+ T cells in the various division classes and the mature effector cells.
Note that effector pool grows and is depleted within the first 20 days, while the proliferating
cells that have not differentiated into effectors are longer lived.

\subsection*{Density-dependent infected cell death}

In the simple coinfection model (\ref{eqs1}), infected cells are assumed to have
constant mortality: $\delta$ for productively infected cells, $T^*$, and
$\mu$ for chronically infected cells, $C^*$, respectively.   In this model
under very effective drug therapy the steady state viral load becomes 
vanishingly small.  However, if the death of
infected cells is made density dependent, as one might expect when infected
cell death is immune-system mediated and immune response is 
a function of the number of infected
cells present, then low steady state viral loads
are possible (Callaway and Perelson, 2002). 

\begin{table}[t]
\begin{center}
\small
\centerline{\textbf{Table 1. Model Parameters}}
\begin{tabular}{|p{40pt}|p{160pt}|p{80pt}|p{60pt}|}
\hline
Parameter & Description & Value & Reference \\
\hline
$a$ & T-cell activation parameter   & variable   &  \\
$\alpha$ & fraction chronically infected   & 0.195   & CP02 \\
$A_{max}$ & pathogen carrying capacity & $10^{8}$ (units/ml) &  \\
$c$ & clearance rate & 23 ($d^{-1}$) &  R99 \\
$d_0$ & death rate, quiescent cells & 0.01 ($d^{-1}$) & DB01 \\
$d$ & death rate, division classes & 0.1 ($d^{-1}$) &  Set  \\
$d_{T_A}$ & death rate, antigen specific & 0.325 ($d^{-1}$) & DB03  \\
$d_{T}$ & death rate, non-specific & 0.01 ($d^{-1}$) & DB01 \\
$\delta$ & death rate, infected cells & 0.7 ($d^{-1}$) & P97  \\
$\delta'$ & density dependent mortality & 0.7863 $d^{-1}$($ml\cdot$cell$^{-1}$)$^\omega$ & CP02  \\
$d_E$ & death rate, effector cells & 0.325 ($d^{-1}$) & DB03  \\
$\epsilon$ & drug efficacy & $0<\epsilon<1$ & Appendix  \\
$\gamma$ &  pathogen clearance rate & $10^{-3}$ ($d^{-1}$) & see text  \\
$k$,$k_1$ & infectivity, single(first) target pool & $8\times10^{-7}$ ($ml\cdot$RNA$^{-1}d^{-1}$) & CP02 \\
$k_A$ & infectivity, Ag-specific target pool & $6\times 10^{-6}$ & set \\
$k_2$ & infectivity, second target pool & $10^{-4}$ & CP02  \\
$K$,$K_8$ & antigen half-saturation for stimulating CD8 cells & variable &   \\
$K_4$ & antigen half-saturation for stimulating CD4 cells & variable &   \\
$\lambda$,$\lambda_1$  & passive T cell source   & $10^{4}$ (cells $/ml/d$)& CP02   \\
$\lambda_2$ & passive T-cell source  & $56$ (cells$/ml/d$)  &  Appendix  \\
$\mu_C$ & mortality, chronically infected  & 0.07   & P97 \\
$\mu'$ & density-dependent mortality    &  0.07863     &  Appendix  \\
$N_0$,$T_0$ & initial quiescent population & 500 cells & see text \\
$N_c$ & burst size, chronically infected & 4.11 RNA copies & CP02 \\
$N_T$ & burst size, productively infected & 100 RNA copies & CP02 \\
$\nu$ & antigen-specific fraction & 0.1 & S98\\
$p_0$ & initial proliferation rate & 1 ($d^{-1}$) &  DB01 \\
$p$ & proliferation rate, classes $1,..,k$ & 2.92 ($d^{-1}$) & DB01\\
$\omega$ & See equation (\ref{omega_eq}) & 0.01 & CP02 \\
$q_1,q_2,q_3$ & HIV production,density dependent & 70,0.28 ($d^{-1}$) & CP02 \\
$q_3$ & HIV production,density dependent & 840 ($d^{-1}$) & Z04 \\
\hline
\end{tabular}
\end{center}
\textbf{CP02}: Callaway and Perelson 2002,
\textbf{DB01}: De Boer \textit{et al.} 2001,
\textbf{DB03}: De Boer \textit{et al.} 2003,
\textbf{P97}: Perelson \textit{et al.} 1997,
\textbf{S98}: Sachsenberg \textit{et al.} 1998,
\textbf{Z04}: Zhang \textit{et al.} 2004.
\end{table}

Here we assume  that the magnitude  of the immune
response against HIV-infected cells, and thus their death-rate,
is an increasing function of the infected cell density (Callaway and Perelson, 2002).  
Holte et al. (2001), fitting data on viral decays in HIV treated children,
suggested that this can be modeled by a power law, that is, by
replacing the constant $\delta$ with a function $\delta^*(T^*)$ that depends on
the total number of infected cells raised to a power $\omega$

\be
\label{omega_eq}
\delta^*(T^*_{total}) = \delta'(T^*_{total})^{\omega}
\ee

and the constant $\mu$ with 

\[
\mu^*(T^*_{total}) = \mu'(T^*_{total})^{\omega}
\]

where $\omega < 1$ and $T^*_{total}=T^*+C$ is the total number of HIV-infected
cells, which represents the stimulus for a cell-mediated response
to HIV-infection.

In the following series of models employing density-dependent
infected cell death, we explore various scenarios for 
CD8+ mediated immune responses to an opportunistic infection.

\subsection*{Simplest case: An HIV infection model with programmed proliferation of CD8+ T cells}

We now combine the basic HIV infection and treatment model 
with CD8+ T cell response to pathogen in the form of
the programmed proliferation model described above. 
Since the effector cells are pathogen specific, 
not HIV specific, the dynamics of the CD8+ T cell population
remains decoupled from the HIV infection model, 
except through their indirect effect on the pathogen.
To the model for programmed proliferation of effector
cells under antigenic stimulation (equations \ref{eqs2})
we add the following model for HIV infection with
density dependent mortality of infected cells:
\begin{subequations}
\label{eqs3}
\begin{eqnarray}
\frac{dT}{dt} & = & \lambda + {f(A)}T - (1- \epsilon)k V T - d_T T \\
\frac{dT^*}{dt} & = & (1- \epsilon)(1-\alpha)k V T - (\delta^*(T^*_{total})T^*  \\
\frac{dC}{dt} & = & \alpha(1-\epsilon)k V T - \mu^*(T^*_{total}) C \\
\frac{dV}{dt} & = & q_1T^* + q_2 C - cV
\end{eqnarray}
\end{subequations}

Target cells, $T$,  are activated by antigen following a type-II functional response,
as in the basic model

\be
 f(A) = a\frac{A}{A + K_4}
\label{eq5}
\ee

and are susceptible to infection by HIV at a rate $k$.
We do not separate out antigen-specific 
CD4+ cells; instead we let the activation parameter $a$ incorporate the
fact that only a fraction of CD4+ cells in that pool may be activated by antigen.

Because the model includes density-dependent cell death, we decouple 
viral production rates from cell death rates, and assume virus is produced
at constant rates $q_1 = N_T \delta = 70$ d$^{-1}$ and
$q_2 = 0.28$ d$^{-1}$, from short-lived and chronically-infected cells,
respectively.
Rate constants and parameters describing the proliferating CD8+ pool remain as
above. A listing of additional fixed parameter values can be found
in Table 1.

\begin{figure}
\centerline{\includegraphics[width=15.25cm,height=15.25cm]{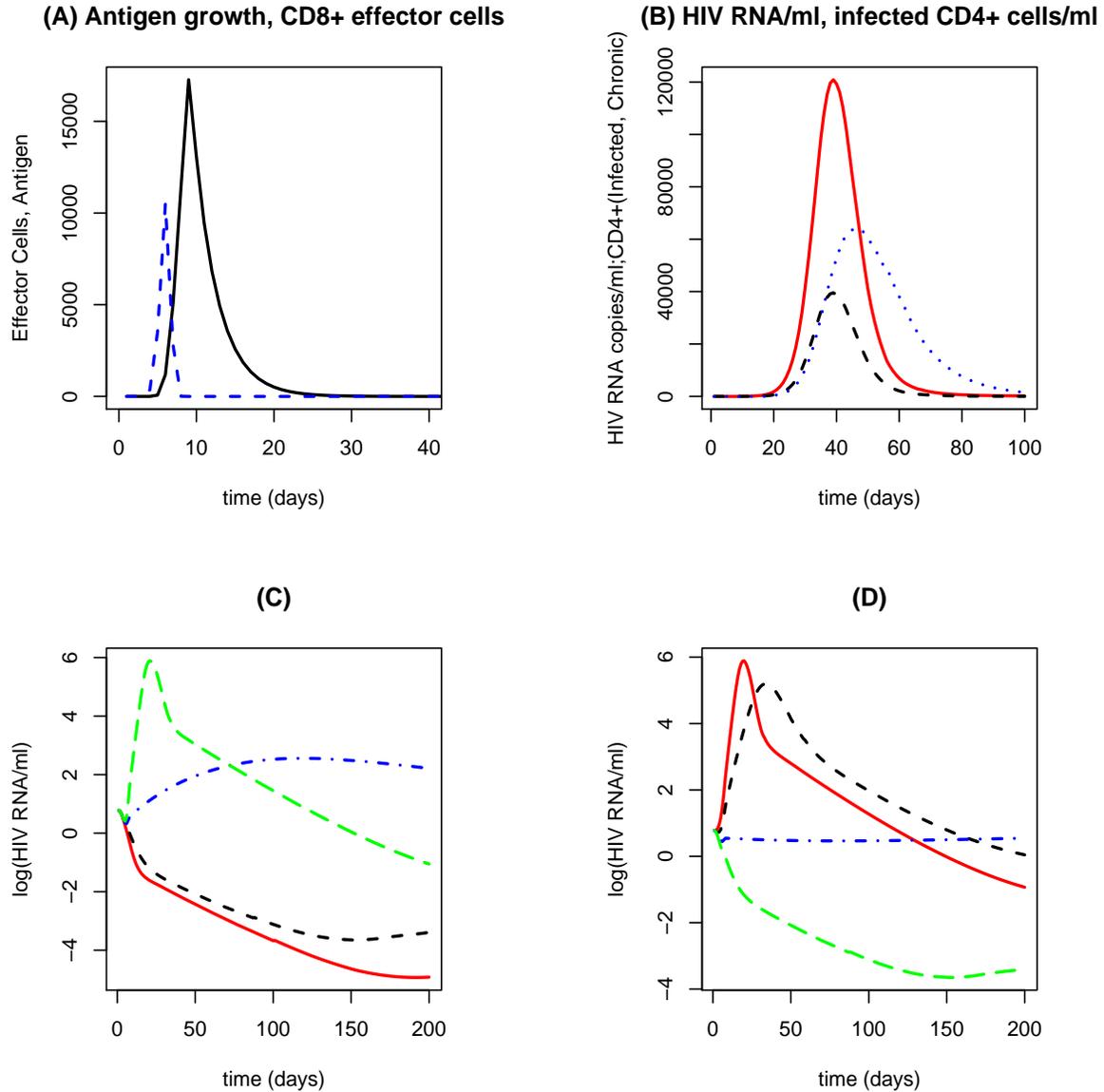}}
\caption{\small  Simulations and sensitivity testing of equations (\ref{eqs3},\ref{fA},\ref{eq5}).
A,Antigen growth (dashed line) and CD8+ effector cells (solid line); B, Viral load (RNA copies/ml, in solid line), productively infected cells ($ml^{-1}$, dashed line), and chronically infected cells ($ml^{-1}$, dotted line);
C, Viral load (copies/ml) from sensitivity testing of  CD4+ T cell
activation parameter $a$, as $a$ assumes the values $a = 0.25,0.50,0.75,1$ (solid, dashed, dash-dotted and long-dashed line, respectively). $r_0=2$ in this example. D, Viral load (copies/ml) from sensitivity testing of 
pathogen growth $K_4$, as $K_4$ assumes the values $K_4 = 1,10^1,10^2,10^3$ d$^{-1}$ (solid, dashed, dash-dotted and long-dashed line, respectively). Note that activation parameter $a$ is held at $0.5$ when not varied. }
\label{Figure 3}
\end{figure}

Figure \ref{Figure 3} shows that this model can generate transient viral load increases. 
Panel~A shows that the  pathogen initially grows and then is 
cleared (dashed line) as the
effector cells (solid line) increase. Pathogen growth and clearance rates are variable, but for this example
we assume $r_0=2$ d$^{-1}$, and clearance rate constant
$\gamma = 1\times 10^{-3}$. The effect of this pathogen growth is to stimulate increased 
infection and viral
production (Fig. \ref{Figure 3}B). 
As drug treatment is assumed, CD4+ levels can be quite
high, leading to an unrealistic number of activated target cells, and excessively large increases
in viral load.
Thus while the model does indeed produce viral transients, there is a tradeoff between the height and the
width of the resulting ``blip". That is, one can obtain an 
appropriately narrow spike in viremia but it is unrealistically
large, 
or one can obtain a blip of appropriate amplitude
(i.e., 100-1000 copies/ml, cf. Di Mascio et al. 2003a) but of unreasonable
duration (upwards of 100 days).  

In our simulations,
parameters associated with steady-state baseline viral load are fixed as
discussed above and in the Appendix, and parameters associated with
pathogen growth and CD4+ activation are considered variable.
Results from sensitivity testing of the model (varying of parameters associated with activation of
target cells in response to pathogen growth and especially the growth
of the pathogen itself) are shown in Figures \ref{Figure 3}C and D.
The model is very sensitive to changes in the CD4+ T cell
activation rate $a$:
between the values $a=0.25$ d$^{-1}$ and $a=0.75$ d$^{-1}$, the model
goes from producing no viral transient to producing an unreasonably
large viral transient (Figure \ref{Figure 3}C).
Variation in the half-saturation of pathogen load required for
CD4+ stimulation, $K_4$, shows the opposite trend: as the
value of $K_4$ and thus the pathogen load required for
CD4+ stimulation increases, resulting viral transients go from being
excessively large (Figure \ref{Figure 3}D, red line), to non-existent.

Variation in $r_0$ shows that beyond a threshold value 
there is relatively small variation in blip peak amplitude with substantial, though biologically reasonable,
variation in pathogen growth rate (results not shown). 

Interestingly, this model exhibits a  delay between the 
peak in opportunistic infection and the peak in viremia (Figure
\ref{Figure 3}B). The delay appears to be related to the time it
takes for the T cell pool to increase and generate a rise in the
numbers of infected cells, $T^*$ and $C$, which then in turn
generate more virus. As one sees (in Figure \ref{Figure 3}C),
increasing the CD4+ T cell activation rate causes the peak in viremia 
to occur earlier. Whether these delays are realistic is unknown,
but because antigen-specific CD4+ cells and ``non-specific" CD4+ cells
are all in the same pool
and the pool has the relatively low mortality appropriate for ``resting" 
cells, the T cell response may not be realistically modeled.

\subsection*{Including separate antigen-specific and non-specific target pools}

In the prior model, the entire CD4+ pool was activated by antigen, resulting in
an overly large target population and correspondingly large viral transients.
To make the model more realistic, we now divide the target (CD4+) cells into non-specific 
and  antigen-specific pools, $T$ and $T_A$, respectively, where
only the antigen-specific CD4+ T cells are activated by antigen.
We couple the model for antigen-induced response of CD8 effectors (\ref{eqs2}) with
the following model for HIV infection of antigen-specific and non-specific CD4
cells: 

\begin{subequations}
\label{eqs6}
\begin{eqnarray}
\hbox{Antigen-specific CD4}\quad
\frac{dT_A}{dt} & = & \nu\lambda + f(A)T_A - (1- \epsilon)k_A V T_A - d_{T_A} T_A\\
\frac{dT_A^*}{dt} & = & (1- \epsilon)k_A V T_A - (\delta^*(\omega))T_A^*  \\
\\
\hbox{non-specific CD4}\quad\frac{dT}{dt} & = & (1-\nu)\lambda - (1- \epsilon)k V T - d_T T \\
\frac{dT^*}{dt} & = & (1- \epsilon)(1-\alpha)k V T - (\delta^*(\omega))T^*  \\
\frac{dC}{dt} & = & \alpha(1-\epsilon)k V T - \mu^*(\omega) C \\
\frac{dV}{dt} & = & q_1 T^* + q_2 C + q_3 T_A^*- cV
\end{eqnarray}
\end{subequations}

where 
$E=\sum_{i=4}^8 N_i$, and $p_0(A)$ is defined in (\ref{fA}) and $f(A)$ defined in (\ref{eq5}).

Antigen-specific CD4+ T cells, $T_A$, are assumed to be small fraction $\nu$  of the
CD4+ pool $T$ (see Table 1), and 
are activated by antigen following a type-II functional response,
as in (\ref{eq5}).   The value of $\nu$ is poorly defined, and may vary widely from
roughly  $10^{-5}$ to $10^{-2}$ (percent activated $0.001\%$ to $0.1\%$, respectively). 
Response to a particular epitope generally occurs at
a frequency of about $10^{-5}$ ($0.001\%$), but a pathogen has many epitopes, thus as many as
10 to 20 responses might occur for a single pathogen, yielding a frequency of $10^{-4}$ (0.01\%).   
However, if one considers levels of activation in HIV-infected versus healthy people, one finds that
in HIV-infected people, roughly $6.5\%$ of CD4+ T cells are
activated, depending on total CD4+ count, versus  about $1\%$ in healthy individuals.
This level of activation might reflect responses to a number of pathogens, or in the case
of immuno-compromised HIV patients, may be a homeostatic response to low CD4+ T cell counts.
Based on these considerations, we choose a frequency $\nu = 0.01$, corresponding to an
activated fraction of about $1\%$.  
(Note that we could easily choose a slightly higher or lower
fraction and obtain similar results by adjusting the activation parameter, $a$ (\ref{eq5})).

Once activated, target cells are subject to infection by HIV at an increased
rate $k_A$. The subsequent increase in viral load results in increased infection of
the non-specific pool, $T$.   Both infected antigen-specific $T_A^*$ and  infected non-specific cells
$T^*$ are cleared in a density dependent fashion.
Rate constants and parameters for the proliferating CD8+ pool remain as above. 
Analysis of SIV infection of resting and infected CD4+ T-cells suggests that
activated, infected CD4+ cells produce on average twelve times more virus
than non-activated infected cells (Zhang et al., 2004).  We thus set $q_3 \doteq 12q_1$
in equations (\ref{eqs6}).

\begin{figure}
\centerline{\includegraphics[width=15.25cm,height=7.6cm]{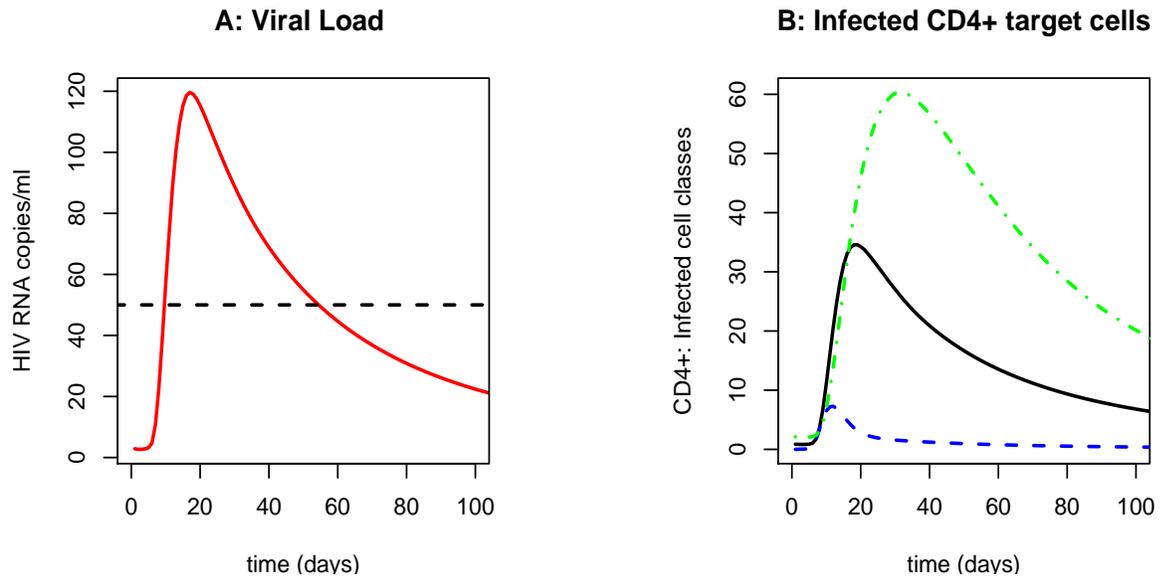}}
\caption{\small  Simulation of equations (\ref{eqs6}) for a patient assumed to be in steady state
with $\bar V\doteq 3$ copies/ml.  
At $t=0$ a pathogen (initial value $A_0=1$), growing at rate $r_0=2.0$ days$^{-1}$, is encountered and eliminated
by an immune response with rate constant $\gamma = 1\times 10^{-3}$. Here $\nu=0.01$, $\epsilon=0.62$,
$a=1.2$, $K_8=1000$, $K_4=100$. A,  Viral load (copies/ml). Dashed line shows limit
of assay detection at $50$ (copies/ml). B,
 Non antigen-specific productively infected (solid line)
 and chronically infected (dashed-dotted line) CD4+ cells, 
 as well as antigen-specific infected CD4+ cells (dashed line),
increased by a factor of 10 to better show dynamics. }
\label{Figure 4}
\end{figure}

Using density-dependent infected cell death, the model can produce realistically low
baseline viral loads under treatment. For this example the baseline viral load is about
3 RNA copies per ml.   Opportunistic infection results in
detectable viral transients (Figure \ref{Figure 4}A), with the stimulus for induced replication arising from
the (small) antigen-specific CD4+ T cell pool, but most of the increase in viral load coming
from infection of antigen non-specific T cells (Figure \ref{Figure 4}B).  The resulting transients have a 
sharp rise-time and
a slow decay, consistent with observation (Di Mascio et al., 2003a). However, 
blip duration is generally too long, upwards of 30 to 90 days, respectively, for transients of 
reasonable amplitude (100-1000 RNA copies per ml). 

\begin{figure}
\centerline{\includegraphics[width=15.25cm,height=15.25cm]{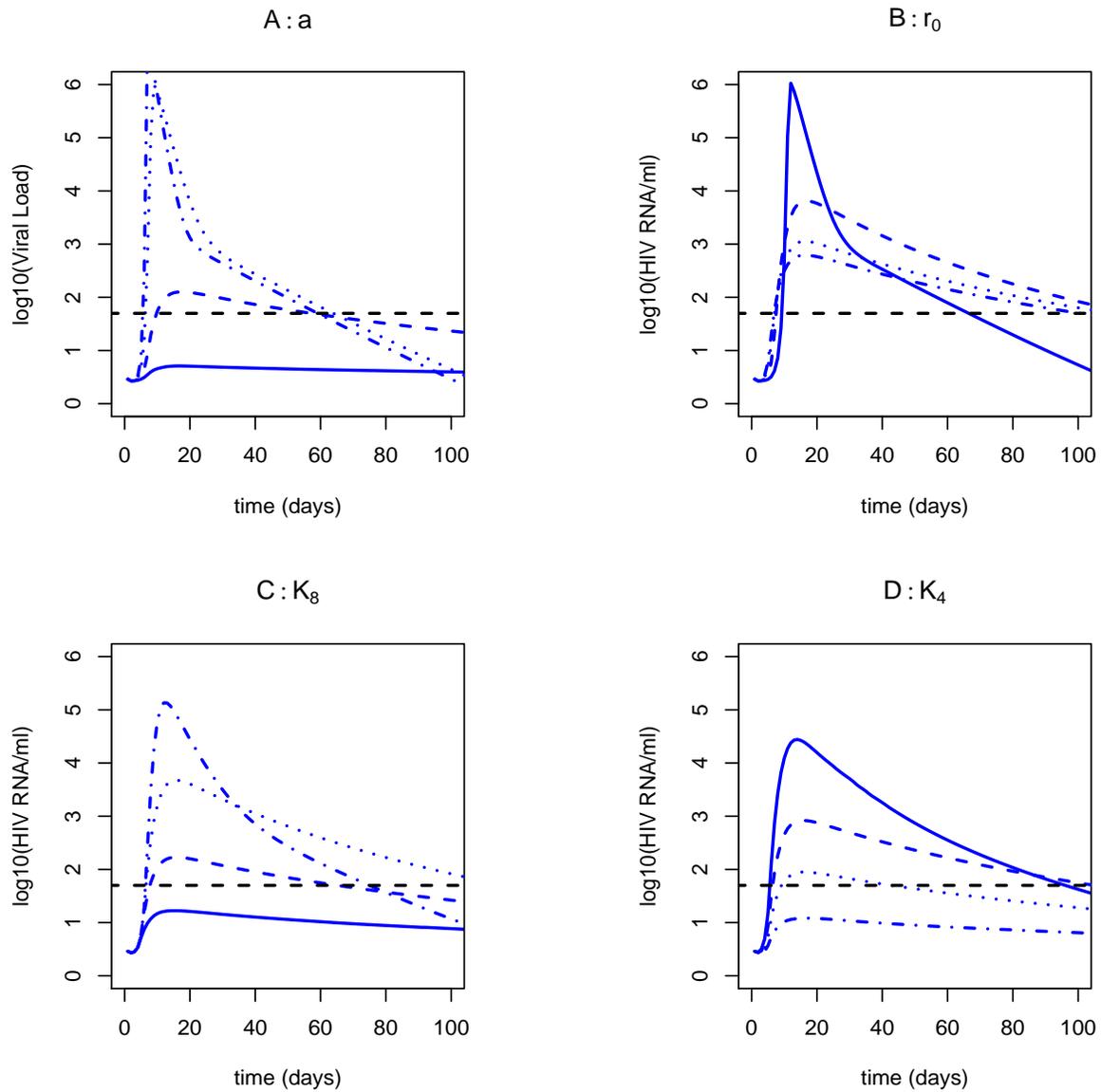}}
\caption{\small  Sensitivity testing, equations (\ref{eqs6}). Unvaried parameters are
as in Figure \ref{Figure 4}.  A, Viral load 
as the activation parameter $a$ varies from $0.5$ (solid line) to $1.5$ (dash-dotted line);
B, Viral load as pathogen growth rate $r_0$ varies from $1$ (solid line) to $4$ (dash-dotted line);
C, Viral load as  $K_8$ (\ref{fA}) varies from $10^2$ (solid line) to $10^5$ (dash-dotted line);
D, Viral load as $K_4$ (\ref{eq5}) varies from $1$ (solid line) to $10^3$ (dash-dotted line).}
\label{Figure 5}
\end{figure}

In addition, the model
is very sensitive to changes in the parameters $a$ and $K_8$, with less
extreme variability in viral load resulting from variation of the pathogen growth 
parameter $r_0$ and the half-saturation parameter $K_4$ (Figure \ref{Figure 5}).
Small variation in the CD4+ cell activation parameter $a$ alone results in a spectrum of behavior from no detectable viremia to excessively large bursts in viral replication (Figure \ref{Figure 5}~A). In addition, though in general infectivity $k$ is
poorly constrained, infectivity for activated antigen specific cells $k_A$ must be
set at levels fifteen times higher than that of quiescent cells to produce detectable viremia, 
and this may not be realistic.

\begin{flushleft}
{\it Effects of parameter variation}
\end{flushleft}
\begin{itemize}
\item  Varying the CD4+ activation parameter, $a$,  results in large changes in peak viremia over a
relatively small range of parameter values, as in Figure \ref{Figure 5}A. Note that blip durations
(i.e., decay times) are too long in all of these examples, and only begin to become shorter with
increased values of $a$ (i.e., greater proportions of the transient arising from infection of
activated, antigen-specific cells).

\item Increasing $r_0$ results in more aggressive pathogen growth and swifter CD8+ immune
response. For lower values of $r_0$, the CD8+ response is delayed, and $f(A)$ (\ref{eq5}) remains elevated longer.
A smaller $r_0$ thus results in a larger amplitude viral transient, with $r_0=1$ days$^{-1}$
resulting in
the largest blip, and successively larger values of $r_0$ resulting in smaller blip sizes
(Figure \ref{Figure 5}B).

\item Increasing the half-saturation constant for CD8+ proliferation, $K_8$ in (\ref{fA}) has a {\em positive} effect
on the sizes of the resulting viral transients, as this allows greater pathogen growth before an immune 
response begins, while increasing the half-saturation for
pathogen growth in the activation function $f(A)$ (\ref{eq5}) has a {\em negative} effect on blip size, as the slope
of the ``activation curve" for target cells becomes gentler (Figures \ref{Figure 5}C,D).

\end{itemize}

A complex but more biologically realistic version of this model would allow proliferation
of the antigen-specific target pool - yielding more activated target cells at the effector stage - 
and might also allow infection of the proliferating
division classes. This would boost the amplitude of the transient and
lower the value of $k_A$ required to produce  detectable viremia, thus addressing two flaws in the
present model (\ref{eqs6}).
We explore both of these options below.

\subsection*{Programmed proliferation of both CD8+ (effector) and antigen-specific CD4+ (target) cells}
CD4+ T cells as well as CD8+ T cells undergo programmed proliferation upon exposure to antigen (Lee et al., 2002). 
These proliferating cells
are more susceptible to HIV infection than resting cells. Thus incorporating CD4 cell
proliferation cascades should affect HIV replication.  We incorporate CD4+ 
proliferation of antigen-specific CD4+ T cells, with each proliferating
division class after the first division assumed activated and thus infectable by HIV-1. 
The model is identical to equations (\ref{eqs6}), with the addition of sets 
of equations for the proliferating $T_0$, $T_i$,... classes,
and the infected $T_1^*$, $T_2^*$,... classes.  

As in prior models we assume CD8+ cells undergo eight divisions, so there are eight division classes
with the assumption of effector status after the fourth division. CD4+ T cells have slower kinetics than
CD8+ T cells, and undergo fewer programmed cycles of proliferation. In the case of
LCMV infection of mice this results in
20-fold lower expansion and a delay by one day until the peak of expansion is reached
(De Boer et al., 2003). To limit
introduction of new parameters, and because little is known about human CD4+ proliferation
rates, we assume the same proliferation rates for CD8+ and CD4+ cells.
Since CD4+ cells
undergo fewer divisions, we assume four CD4+ division classes as opposed to eight,
and further assume that they are activated by pathogen and infectable after the first division, and
that they suffer elevated mortality $\delta_A$ when activated.
In addition, we allow 
continued proliferation (division) of infected division classes, which divide to produce
infected cells.
The initial number of naive cells in both the CD8+ (i.e., $N_0$) 
and CD4+ (i.e., $T_0$) pools is set at 500 cells. 
To the model for programmed proliferation of CD8 effectors under antigenic stimulation
we add the following equations for proliferation and infection of antigen-specific CD4
cells and the passive infection of the nonspecific CD4 pool:

\begin{subequations}
\label{eqs8}
\begin{eqnarray}
\hbox{Antigen-specific CD4} \quad\frac{dT_0}{dt} & = &   -( p_0(A) + d_0)T_0 \\
\frac{dT_1}{dt} &  = & 2 p_0(A)T_0 - (p + d)T_1 - (1-\epsilon)k_A V T_1 \\
\frac{dT_{i}}{dt} & = &\begin{cases} 2p T_{i-1}- (p + d)T_{i}- (1-\epsilon)k_A V T_{i},  i=2,3 \cr
                      2p T_i - (1- \epsilon)k_A V T_{i+1} - d_{T_A} T_{i+1}, i=4  \cr \end{cases} \\
\frac{dT^*_1}{dt} &  = & (1-\epsilon)k_A V T_1  - (\delta^*(\omega) + p)T^*_1\\
\frac{dT^*_{i}}{dt} & = &\begin{cases} (1-\epsilon)k_A V T_{i} + 2p T^*_{i-1} -(\delta^*(\omega) + p)T^*_{i}, i=2,3 \cr
                         (1- \epsilon)k_A V T_i  + 2p T^*_{i-1} - (\delta^*(\omega))T_i^*, i=4 \end{cases}  \\
\hbox{Non-specific CD4} \quad \frac{dT}{dt} & = & \lambda  - (1- \epsilon)k V T - d_T T \\
\frac{dT^*}{dt} & = & (1-\alpha)(1- \epsilon)k V T - (\delta^*(\omega))T^*  \\
\frac{dC}{dt} & = & \alpha(1-\epsilon)k V T - \mu^*(\omega) C \\
\frac{dV}{dt} & = & q_1 T^* + q_2 C + q_3\sum T_i^*- cV
\end{eqnarray}

\end{subequations}

where  $E=\sum_{i=4}^8 N_i$ and $p_0(A)$ is defined in equation (\ref{fA}).

\begin{figure}
\centerline{\includegraphics[width=15.25cm,height=7.0cm]{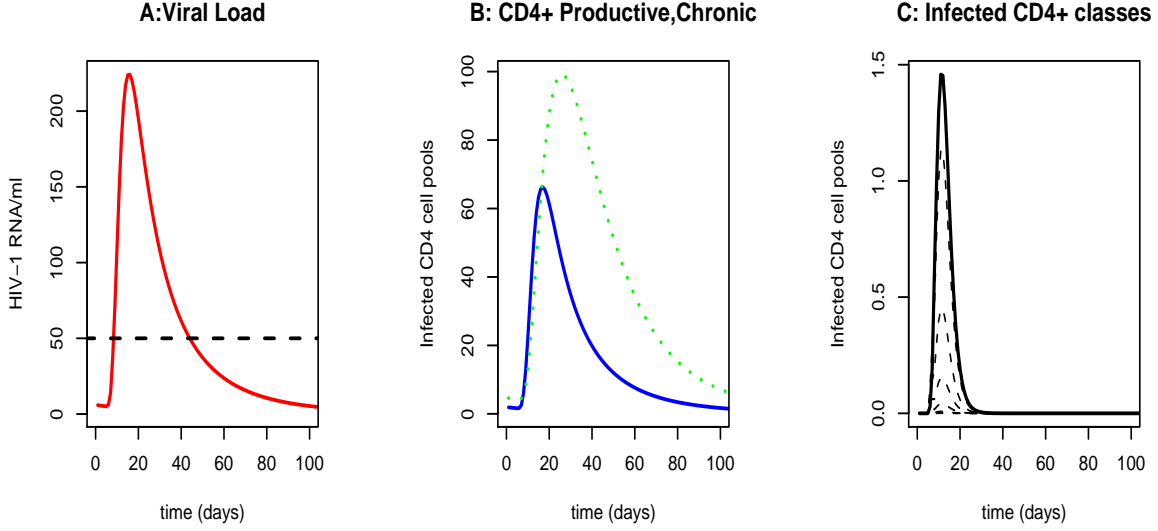}}
\caption{\small  Simulation of equations (\ref{eqs8}).  Again a patient is assumed to be at steady
state ($\bar V = 3$ copies/ml) 
when at $t=0$ a growing pathogen is encountered. Here, $k_A = 4\times 10^{-6}$ml/d, $r_0=2.0$ d$^{-1}$
and $\gamma = 1\times 10^{-3}$ ml/d.
Other parameters are as described in text and shown in Table 1.
A, Viral load (copies/ml); dashed line shows limit of assay detection at 50 (copies/ml). 
B, Total productively (solid line) and chronically infected CD4+ cells (dashed line), 
C, Infected antigen-specific CD4+ effectors, solid line, plotted with infected antigen specific CD4+ division classes, thin dashed line. }
\label{Figure 6}
\end{figure}

\begin{figure}
\centerline{\includegraphics[width=15.25cm,height=7.0cm]{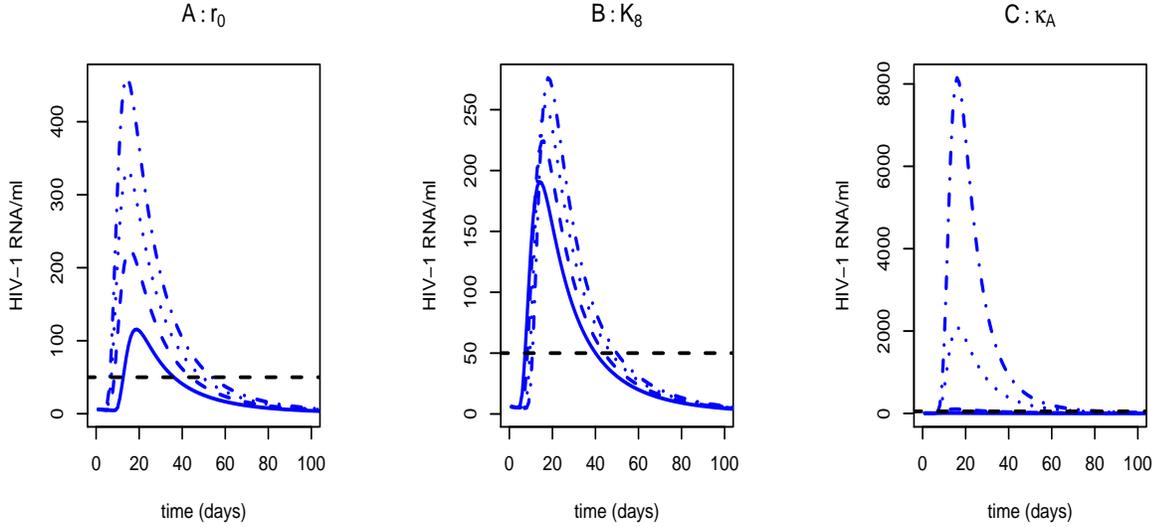}}
\caption{\small  Sensitivity testing for equations (\ref{eqs8}).  Each panel shows viral load (copies/ml)
on a linear scale.
A, Viral load as pathogen growth parameter varies, $r_0=1,2,3,4$ d$^{-1}$
in solid, dashed, dotted, and dot-dash line, respectively;
B, Viral load as $K_8$ (\ref{fA}) varies from $10^2$ (solid line) to $10^5$ (dash-dotted line);
C, Viral load due to variation in antigen-specific, activated cell infectivity
 $k_A = 8\times 10^{-7}$ 
({\em cf.} Callaway and Perelson, 2002), $1.6\times 10^{-6}$, $3.2\times 10^{-6}$,
$6.4\times 10^{-6}$, $8.0\times 10^{-6}$ ($ml\cdot$RNA$^{-1}d^{-1}$); solid, dashed, dotted, and dot-dash line, respectively.}
\label{Figure 7}
\end{figure}

Figures \ref{Figure 6} and \ref{Figure 7} show simulation results and sensitivity testing
for equations (\ref{eqs8}).   Figure \ref{Figure 6}A shows viral load,  panel B shows productively and
chronically infected cells (solid and dashed line, respectively), and panel C shows infected antigen-specific CD4+ effectors (bold solid line) and division classes (thin dashed line) for a representative parameter
set ($r_0 = 2$ d$^{-1}$, $K_8 = 10^3$, and $k_A=6\times10^{-6}$).  The rise in infected and chronically infected cells from the
non-specific pool is fairly small, but most of the new replication in the transient is coming from infection of the
even smaller activated antigen-specific pool. The timing of the peak in new productively and chronically infected
cells from the non-specific pool is too late to account for the bulk of the viremia in the transient.

The model is fairly robust, producing viral transients of biologically reasonable amplitude and
duration over a range of parameter values.  In Figure \ref{Figure 7} we test the model for reasonable
variation in three parameters critical to pathogen growth and viral replication. Results for all three
tests are shown in linear scale, as the variation in peak viremia spanned only one order of magnitude.
Panel A shows variation in viral load (copies/ml) as $r_0$ is varied from $r_0=1$ d$^{-1}$ to 
$r_0=4$ d$^{-1}$ , while parameters
$K_8$ and $k_A$ are held at the values listed above, and all other parameters are held constant
as listed in Table 1. Unlike the
prior model (equations \ref{eqs6} and Figures \ref{Figure 4} and \ref{Figure 5}), in this case peak viral load is an 
increasing function of $r_0$, though the increases in VL with each increase in $r_0$ are moderate.  In panel
B, $r_0$ is again set to $r_0=2$ d$^{-1}$, and the parameter $K_8$ is varied from $10^2$ (solid line) to $10^5$ (dash-dotted line),
resulting once again in moderate increases in peak viral load and blip duration.  Variations in the infectivity of the
antigen-specific cells $k_A$  over a biologically reasonable range of values also results in moderate and
acceptable increases in peak blip viremia (Figure \ref{Figure 7}C).

\section*{Models with heterogeneous target pools}

Another intuitive explanation for the presence of intermittent viremia is to assume that
different cell types, or cell populations, might have different drug penetrances, perhaps due to
different levels of expression of P-glycoproteins that have the potential to pump drugs
 out of cells (Kim {\it et al.}, 1998). Having cell populations with
reduced drug penetration may produce episodes of transient viremia in the
presence of opportunistic infection. Indeed, there is evidence from {\it in vitro} studies of
heterogeneity in intracellular drug concentrations ({\it cf.} Kim {\it et al.}, 1998;
Perno {\it et al.}, 1998; Puddu {\it et al.}, 1999).  In the following model
we include two co-circulating populations of target cells with differing drug penetration,
wherein drug efficacy is reduced in one cellular population. 
This is, in addition, a  simple means of producing low steady-state viral loads 
in the presence of drug therapy which does not invoke density dependent
mortality of infected cells ({\em cf.} Callaway and Perelson, 2002).

\begin{figure}
\centerline{\includegraphics[width=15.25cm,height=15.25cm]{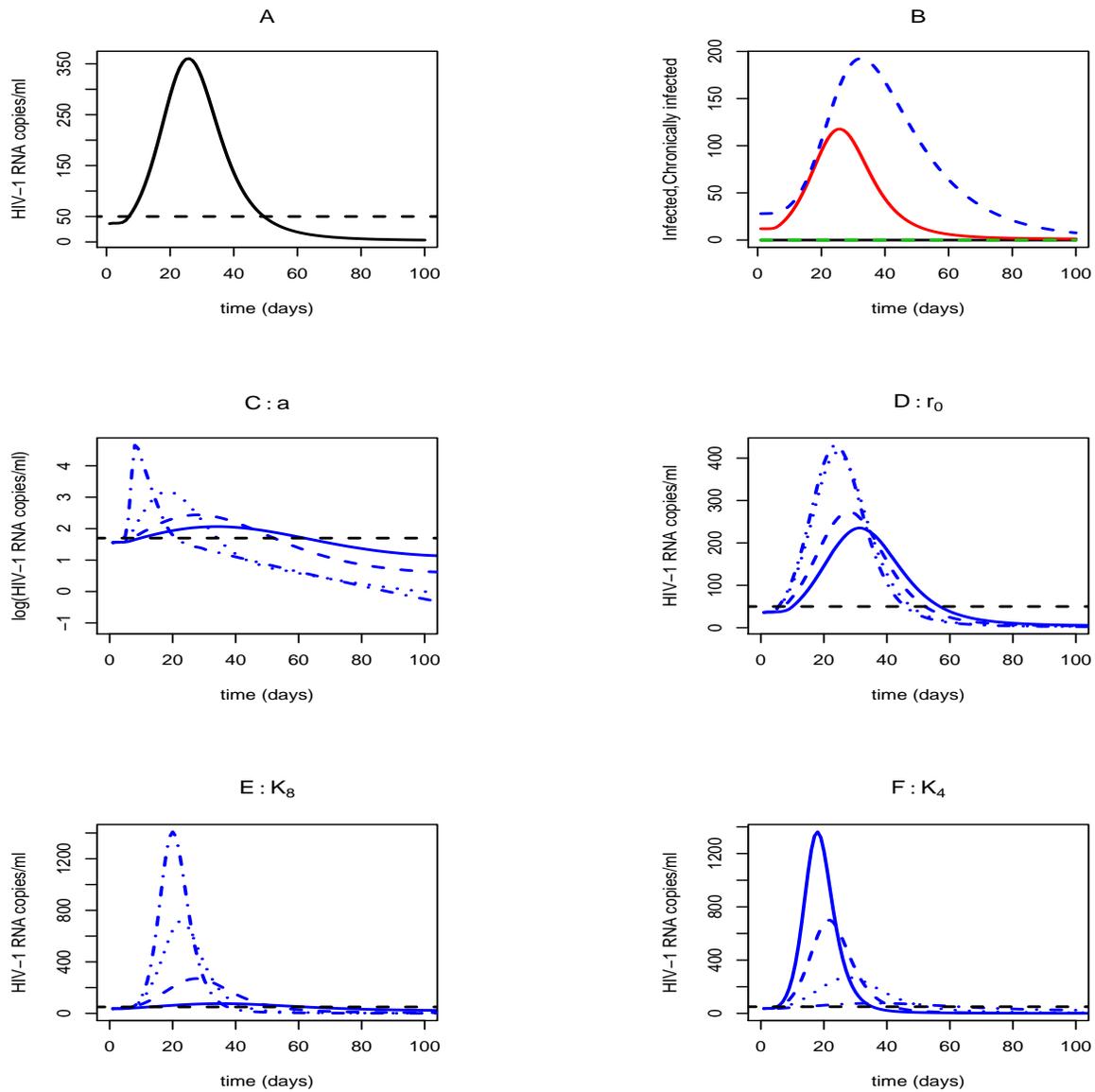}}
\caption{\small  Simulation of equations (\ref{eqs9}). Here, $k_1 = 8\times 10^{-7}$,
$k_2 = 1 \times 10^{-5}$ and other parameters are as described in text and shown in Table 1.
A, viral load (solid line, in RNA copies/ml) assuming $r_0=2.0$ and $\gamma = 1\times 10^{-3}$. 
B, Productively and chronically infected CD4+ cells, dashed and solid lines respectively. 
Note that these cells are contributed only by the cellular pool for which drug efficacy is
assumed to be imperfect. C, Viral load ($log_{10}$) as the activation parameter $a$ varies from $0.01$ (solid line) to $1$ (dash-dotted line); D, Viral load (linear scale)as pathogen growth rate $r_0$ varies from $1 d^{-1}$ (solid line) to $4$ (dash-dotted line);
E, Viral load as $K_8$ (\ref{fA}) varies from $10^2$ (solid line) to $10^5$ (dash-dotted line);
F, Viral load as $K_4$ (\ref{eq5}) varies from $10^1$ (solid line) to $10^4$ (dash-dotted line).}
\label{Figure8}
\end{figure}

The model again includes expansion of antigen-specific CD8+ cells, with the
effector class $E$ as the final division class. Effector cells have
the same mortality as mature CD4+ T-cell classes, but remain decoupled from the
two co-circulating CD4+ classes $T_1$ and $T_2$. Target cells do not proliferate
from the naive state but are assumed to have passive sources
 $\lambda_1$ and $\lambda_2$ (Callaway and Perelson, 2002).
The model assumes differential
drug efficacy in the co-circulating target cell populations: 
in one population ($T_1$), drug efficacy is $\epsilon$, and in the second population
($T_2$), drug efficacy is  $f\epsilon$, where efficacy has been reduced by a factor $f< 1$.
Finally, the model includes two separate chronically infected cell pools, $C^*$, derived from infected
non antigen-specific T-cell pools, $T_1^*$ and $T_2^*$.  Again, the model for programmed proliferation
of antigen-specific CD8 effector cells is as in (\ref{eqs2}).

\begin{subequations}
\label{eqs9}
\begin{eqnarray}
\frac{dT_1}{dt} & = & \lambda_1 + f(A)T_1 - (1- \epsilon)k_A V T - \delta T \\
\frac{dT_2}{dt} & = & \lambda_2 + f(A)T_2 - (1- f\epsilon)k_2 V T - \delta T \\
\frac{dT_1^*}{dt} & = & (1-\epsilon)(1-\alpha)k V T_1 - \delta T_1^*  \\
\frac{dT_2^*}{dt} & = & (1-f\epsilon)(1-\alpha)k V T_2 - \delta T_2^*  \\
\frac{dC_1}{dt} & = & \alpha(1-\epsilon)k V T_1 - \mu C_1 \\
\frac{dC_2}{dt} & = & \alpha(1-f\epsilon)k V T_2 - \mu C_2 \\
\frac{dV}{dt} & = & N_T \delta (T_1^* + T_2^*) + N_c \mu( C_1 + C_2) - cV
\end{eqnarray}
\end{subequations}

where again $E = \sum_{i=4}^8 N_i$,
$ p_0(A)$ is as in equation (\ref{fA}), and $f(A)$ is as in equation
(\ref{eq5}).
CD8 effector cells are activated by antigen following a type-II functional response,
as in the basic model  (\ref{eqs3}). 
As in equations (\ref{eqs2}), CD4+ cells in both pools are activated 
following a type-II functional response, with activation parameter $a$.
Other parameters are as defined above and
in Table 1, with the exception of parameters specific to a given target cell pool, which
are given subscripts appropriate to that pool. Infection rates $k_A$ and $k_2$
and cell production rates $\lambda_1$ and $\lambda_2$
are thus specific to target cell pools $T_1$ and $T_2$, respectively.  As in Callaway and Perelson (2002),
we do not introduce different infected cell death rate constants $\delta$ and $\mu$ for
the differing target cell types.   Assuming that $\epsilon =1$ in population 1, the steady state viral
load is

\be
\label{ss2}
\bar V(\epsilon=1)= \frac{\lambda_2}{c} [(1-\alpha)N_T + \alpha N_c] - \frac{d_2}{(1-f)k_2}.
\ee

Further assuming that $\bar V = 100$ when $f=0$, we set 
$\lambda_2=57$ cells ($\textrm{ml}^{-1}$ $\textrm{d}^{-1}$, see Appendix)
and $k_2  = 10^{-4}$ ml $\textrm{copies}^{-1}$ $\textrm{d}^{-1}$, 
again as in Callaway and Perelson (2002).  

In Figure \ref{Figure8}
are shown simulations and sensitivity testing of (\ref{eqs9}).
Panel A shows viral dynamics assuming a pathogen with growth rate $r_0=2$; viral load is
shown in solid line and the threshold for detection [50 copies/ml] is shown dashed.
In Figure \ref{Figure8}~B is shown the rise in productively and chronically infected
cells for both target cell populations. Infected cells for target cell population 1
do not rise at all, so all new viral replication and thus the entire viral transient
comes from newly infected cells in population 2.  Finally,
the above example shows that opportunistic infection will produce viral
transients of typical size and duration in numerous models with suitable
mechanisms for generating robust low steady-state viral loads.

\section*{Discussion}
Models of HIV infection have generally not included detailed models of immune responses.
When immune responses are included in such models, they tend to be very stylized,
phenomenological rather than mechanistic, and with simple predator-prey type dynamics (cf., Nowak
and Bangham, 1996; Wodarz et al., 2000; Wodarz and Nowak, 2000).
Recent experimental work has shown that both the CD4+ and CD8+ classes of T cells undergo
bursts of proliferation on encounter with antigen, and that programmed proliferative 
responses continue even if antigen is removed (Badovinac et al., 2002).  This is a feature
 that has not previously been included in HIV models,
and one that is especially important to consider when simulating the effects of coinfection  - infection with a
growing pathogen - on chronically infected HIV patients.

Here we have built a set of models of increasing realism that captures the proliferative response
of both CD8+ and CD4+ T cells, and the dynamics of chronic HIV infection in the presence of
opportunistic infection.  
In order to tie our models to data we have focused on explaining the 
rare occurrences of transient episodes of viremia, or viral blips, that are seen in patients
treated with potent antiretroviral therapy.   Blip frequency is independent of time on treatment,
and if viral loads were to continue to fall towards extinction as they do in most standard models,  one
would expect the occurrence of detectable viral transients 
to become rarer with increased time on treatment, something which is not observed (Di Mascio et al., 2003a).  
Furthermore, chronically infected patients are usually on therapy and their viral loads are
generally suppressed below, or close to, the level of detection by standard assay. For 
these reasons  we  focused our analysis on HIV models that allow 
robust low steady state viral loads (Callaway and Perelson, 2002).\\

{\it Simulation results} \\
Starting from a very simple model of T cell and HIV dynamics under drug 
therapy (equations \ref{eqs1}), which when challenged with a growing pathogen
only produced a gradual increase in viral load rather than a blip (Figure 1C)
we built a family of new models, at each step adding more biological realism and complexity.
The results from each successive interim model guided us towards the creation of  
biologically realistic and robust end-models, which are capable of producing viral transients of
``typical" duration and amplitude, under a biologically reasonable parameter regime.

Addition of a proliferating effector pool and density dependent infected cell death 
to the initial model results in  a model which can produce viral transients, but with an inherent trade-off between 
blip duration and amplitude  (equations \ref{eqs3}).
The model either produces a blip of appropriate duration, but of amplitude several orders of
magnitude larger than desired, or of appropriate amplitude, and upwards of 100 days duration
(Figure \ref{Figure 3}) depending on the choice of parameters.  This appears to be due to the fact that
in the model the entire pool of CD4+ cells is activated by antigen, resulting in a large spike in viral replication.
In addition, this large target pool has the low mortality assumed for ``resting'' cells, which affects
the duration of the resulting blips.

We then separate the target pool into antigen specific and non-specific CD4+ T cells, and
allow activation by antigen - and thus increased susceptibility to infection - of the 
antigen-specific pool (equations \ref{eqs6}).  Simulations of this model can produce blips with a sharp
rise-time and slow decay, consistent with observation, and  of ``typical" amplitude as described above.
However, even the smaller ($<100$ RNA copies/ml) blips are slightly too long in duration ($~ 30$ days)
(Figure \ref{Figure 4}A), and the model is very sensitive to changes in the activation
parameter $a$ and the half-saturation constant $K_8$, both of which are difficult to quantify
via experiment or observation.

Allowing the antigen-specific target cell pool to proliferate upon exposure to antigen 
is a natural next step, which eliminates the biologically ambiguous activation parameter, $a$, and produces a model
which generates ``blips'' of reasonable amplitude and duration over a range of
parameter values (equations \ref{eqs8}; Figure \ref{Figure 7})

A  model that includes cell populations with different drug susceptibilities 
(equations \ref{eqs9}) generates robust low steady state
viral loads without having to invoke density--dependent cell death and its attendant parameters
(cf. Callaway and Perelson, 2002). This model
when coupled with a CD8+ T cell model of the response to a replicating antigen, 
can also produce viral transients of appropriate
amplitude and duration. In this case it is interesting to note that the viral
transient arises entirely from the second compartment with reduced drug penetration; thus
the size and duration of such transients might scale with the size of the compartment
from which they arise.

{\it Limitations}\\
The models presented here, despite their increasing complexity, remain simplifications of an
in vivo system.   In order to limit the size of our models and
the number of parameters therein,
we have not included HIV-specific CD8+ and CD4+ T cells.
In addition, we have focused entirely on a T-cell mediated immunity, and have not included
antibody (B-cell) responses.  

The clonal expansion and survival of CD8+ T cells may depend on the activity
and ``help" of CD4+ T cells, but mostly during formation of memory and during antigen rechallenge (Bevan, 2004). 
Thus {\em primary} CD8+ T cell responses are undiminished in a CD4+ T cell deficient system: 
when antigen alone invokes an overwhelming
inflammatory response, as in many infections, primary CD8+ T cell responses are entirely independent of the CD4+ T cell response (Bevan, 2004). Yet CD8+ T cell memory during antigen {\em rechallenge} does not function properly in a CD4+ T  cell deficient system.
Furthermore, there is some evidence that frequency of viral transients is statistically correlated with low CD4+ counts (Di Mascio et al., 2003a).  It would
be worth exploring whether, in a CD4+ T cell deficient environment, insufficient CD8+ T cell responses to chronic immune rechallenge contribute to episodes of transient viremia.
While we do not include CD4+ ``help", or indeed, CD8+ T cell
memory in our system, these features would be necessary for any biologically realistic simulation
of an immune system encountering pathogens in a stochastic fashion.   We have omitted these features here as we
have not modeled antigen rechallenge and immune memory.  Our future research interests include stochastic exploration of immune activation during random encounters (and re-encounters) with pathogen, and thus will invoke immune memory.

{\it Conclusions}\\
We have shown that episodes of transient viremia in otherwise well--suppressed, chronically 
infected HIV patients under drug therapy may be triggered by randomly--
occurring opportunistic infections, which cause
a rise in activated target cells, and thus transient bursts of infection and resultant viremia.  Simple 
models for coinfection, which incorporate stylized, predator-prey 
type immune dynamics, cannot account for the
occurrence of viral transients.  However, if biologically realistic T cell dynamics that include
programmed proliferation and contraction of the CD8+ T cell are incorporated into the model, along with
a mechanism for producing robust low steady-state viral loads under HAART, then we have shown
that opportunistic infection can serve as
a ``forcing function'' for transient episodes of viral replication and can explain the observation of viral
blips in otherwise well-suppressed HIV patients.  This is just one plausible mechanism among many for the
generation of blips;  the list of 
possible causes also includes missed drug doses, activation of latently infected cells and release of virus 
from tissue reservoirs. It remains to be shown that these other mechanisms also work in a model and produce transients
of appropriate amplitude and duration.  

\section*{Acknowledgements}
This research was supported by the Department of Energy under contract W-7405-ENG-36 
and by NIH grants AI28433 and RR06555 (ASP).  We thank Michele Di Mascio for the data
used in Figure 1.  We thank Rob DeBoer for many useful
discussions on the topic of T-cell proliferation, and the Cornell EEB eco-theory lunch-bunch, 
in particular Stephen Ellner, for comments which greatly improved this paper.
\pagebreak
\section*{Appendix}
All simulations were written
in the \textbf{R} programming language (version 1.9) and run on a 
Windows XP platform, using the \textbf{odesolve} package. 
 
Steady state values were obtained numerically by setting fixed parameter values
and running the model until 
the simulation reached steady state.  Steady state values for each variable were noted and
used as initial conditions for simulations of the effect of concurrent infection.

{\bf Parameter values}:\\
$\epsilon$: for the chronically infected cell model (\ref{eqs1}), we seek a steady
state viral load $\bar V$ of around 50 copies/ml.  Solving the expression for steady state
viral load (\ref{ss1}) for $\epsilon$, we obtain:
$$ 
\epsilon =  1 - \frac{dc/\kappa}{\lambda\left[(1-\alpha)N_T + \alpha N_C\right] -c \bar V}.
$$
Substituting in the appropriate parameters, one obtains a required drug efficacy of
about $0.6458$ for a steady state viral load of $\bar V\doteq50$.  This efficacy 
is close to the critical efficacy for the models presented here, given our parameters.

$\lambda_2$: In the two-compartment co-circulating cell model (\ref{eqs9}),
we invoke the expression for steady-state viral load, assuming $\epsilon=1$ in the
first cellular pool (\ref{ss2}). Solving for the parameter $\lambda_2$:
$$
\lambda_2=c\frac{\bar V\kappa_2(1-f) + d_2}{\kappa_2(1-f)[(1-\alpha)N_T + \alpha N_C]}
$$
Further assuming that $\bar V = 100$ when $f=0$
and substituting in appropriate parameter values gives a value $\lambda_2\doteq57$.

$\mu'$:  Observing that the ratio $\delta:\mu = 10$ for the death rates of
productively versus chronically infected cells (Callaway and Perelson, 2002) we
retain the same ratio of death rates under the assumption of density dependent
mortality for both cellular pools. Thus, if $\delta'=0.7863$ $d^{-1}$($ml\cdot$cell$^{-1}$)$^\omega$,
then $\mu'=0.07863$ $d^{-1}$($ml\cdot$cell$^{-1}$)$^\omega$.

\pagebreak

\end{document}